\documentclass[onecolumn,11pt,a4]{revtex4}
\usepackage{epsfig,graphicx,float,pst-all}
\usepackage{amssymb}
\usepackage{mathrsfs}
\usepackage{amsmath}
\usepackage{url}
\usepackage{esvect}
\usepackage[T1]{fontenc}
\usepackage{textcomp}
\usepackage{gensymb}
\usepackage{url}
\usepackage{multirow,rotating}
\usepackage{setspace}

\newcommand{\be}{\begin{equation}} 
\newcommand{\ee}{\end{equation}}
\newcommand{\bea}{\begin{eqnarray}} 
\newcommand{\eea}{\end{eqnarray}}
\newcommand{\bc}{\begin{center}}
\newcommand{\ec}{\end{center}}

\begin{document}

\title{ 
Electric multipole response of the halo nucleus $^6$He}
\author{Jagjit Singh}
\email{jsingh@pd.infn.it}
\author{L. Fortunato}
\author{A. Vitturi}
\affiliation{Dipartimento di Fisica e Astronomia \textquotedblleft G.Galilei\textquotedblright\\}
\affiliation{INFN-Sezione di Padova, via Marzolo 8, I-35131 Padova, Italy}
\author{R. Chatterjee}
\affiliation{Department of Physics, Indian Institute of Technology, Roorkee 247 667, India}

\vspace{1cm}

\date{\today}

\begin{abstract}
The role of different continuum components in the weakly-bound nucleus $^6$He is studied by coupling unbound spd-waves of $^5$He by means of simple pairing contact-delta interaction. 
The results of our previous investigations in a model space containing only p-waves, showed the collective nature of the ground state and allowed the calculation of the electric quadrupole transitions. 
We extend this simple model by including also sd-continuum neutron states and we investigate the electric monopole, dipole and octupole response of the system for transitions to the continuum, discussing 
the contribution of different configurations.
\end{abstract}

\maketitle

\section{Introduction}
\label{intro}
\begin{figure*}
\centering
\vspace*{0.2cm}
\includegraphics[width=0.99\textwidth, clip=]{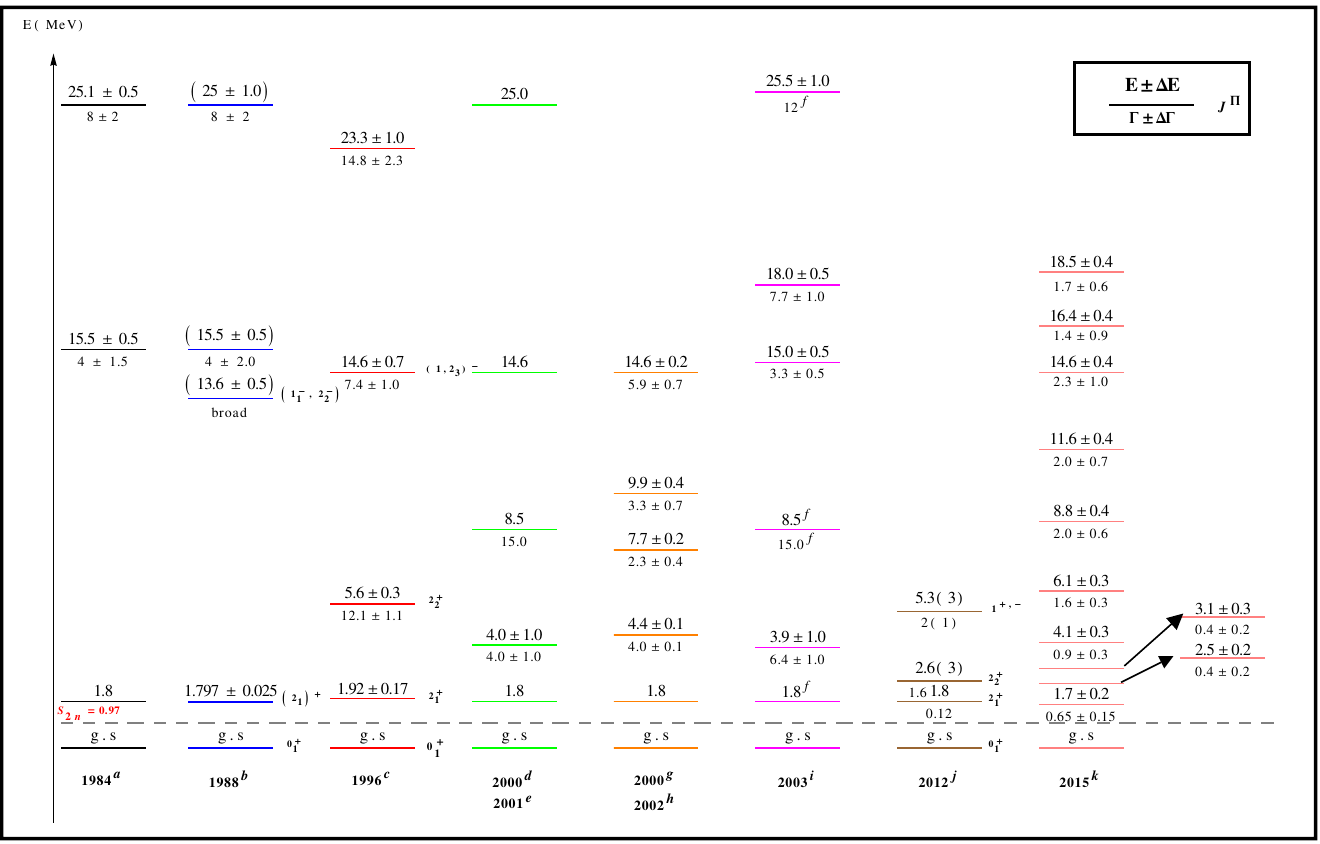}
\caption{(Color online) Experimental spectroscopy of $^6$He.
$^a$Reference \cite{Brady}, \textit{(n,p)} at $60$ MeV/nucleon.
$^b$Reference \cite{Ajzen}.
$^c$Reference \cite{Janeke}, ($^7$Li, $^7$Be) at $50$ MeV/nucleon.
$^{d,e}$Reference \cite{Nakay,Nakay2}, ($^7$Li, $^7$Be) at $65$ MeV/nucleon.
$^{g,h}$Reference \cite{Naka,Naka2}, (\textit{t}, $^3$He) at $112$ MeV/nucleon.
$^i$Reference \cite{Akimune}, ($^7$Li, $^7$Be \textit{t}) at $65$ MeV/nucleon.
$^j$Reference \cite{Moug}, ($^8$He, \textit{t}) at $15.4$ MeV/nucleon.
$^k$Reference \cite{Povo}, ($\alpha$, \textit{p}$\alpha$) at $67.2$ MeV/nucleon.
$^\textit{f}$Fixed in fits.}
\label{Expts}
\end{figure*}

Due to the recent developments in the radioactive beam facilities around the world, it is possible to explore new phenomena in unstable nuclei far from line of stability.
In the vicinity of neutron dripline, a neutron halo is the most intresting phenomena found in some  unstable nuclei \cite{Tanh1}.
Typical two-neutron halo nuclei are $^6$He \cite{Tanh2} (system under study), $^{11}$Li \cite{Tanh1}, $^{14}$Be \cite{Tanh3} and recently observed $^{22}$C \cite{Tank}. 
These two-neutron halo nuclei are referred as Borromean nuclei \cite{Zhu}, when there is no bound state between a valence neutron and a core nucleus.
Borromean nuclei typically have small two-neutron separation energy $(S_{2n})$. 
Due to diminishing half-lives and narrow production cross sections, the experimental analysis of these dripline systems is a challenging issue.
In these weakly-bound nuclear systems, the properties of the continuum states become progressively more and more fundamental to the nuclear structure and reactions.
On the theoretical side the treatment of low breakup thresholds, reponsible for strong coupling of bound and continuum states is the challenging issue. 
A low breakup threshold introduces many new features such as large spatial density distribution \cite{Tanh1, Tanh2}, evolution of new magic numbers \cite{Ost}, a narrow momentum distribution \cite{Kob} and 
at lower excitation energies strong concentration of electric dipole strength \cite{Fuk, Aum, Nak, Vitt4} in these systems.
In this paper we study the electric multipole response of the well established halo nucleus $^6$He. 
Experimentally the higher excited states of $^6$He are still under discussion and the features of resonance states are still not very clear.
In the eighties, the $J_{\pi}=0^+$ ground state and first excited $J_{\pi}=2^+$  state at energy $1.797$ MeV have been confirmed in numerous reactions \cite{Brady, Ajzen}.
In late nineties, the $^6$Li($^7$Li,$^7$Be)$^6$He charge-exchange reaction has been studied at E$(^7$Li$) = 350$ MeV and transitions to the known $J_{\pi}=0^+$ ground state and 
the $J_{\pi}=2^+$ state at $E_x = 0.0$ and $1.8$ MeV (weak) and three strong and broad resonances at $E_x \approx 5.6, 14.6$ and $23.3$ MeV have been observed \cite{Janeke}.
The strong resonances at $\sim5.6$ MeV and $\sim14.6$ MeV are interpreted as $2^+$ and ($1, 2$)$^-$ resonances, respectively.
Proton-neutron exchange reactions between two fast colliding nuclei produced resonant-like structures around 4 Mev \cite{Nakay} of width $\Gamma\sim4$ MeV, as well as asymmetric bump at$\sim5$ MeV \cite{Naka}, 
and these structures are explained as dipole excitations compatible with oscillations of positively charged $^4$He core against halo neutrons. Different mechanisms have also been proposed to explain this mode 
and this phenomenon is still under debate.
More recently, the two-neutron transfer reaction p($^8$He,t) at the SPIRAL facility at 15.4 $A$MeV (GANIL, Caen), populated a much narrower
$2^+$ ($\Gamma=1.6$ MeV) state and a $J=1$ resonance ($\Gamma\sim2$ MeV) of unassigned parity at energies $2.6$ and $5.3$ MeV respectively \cite{Moug}.
It is worthwhile to mention that a very new reaction $^3$H($\alpha$,$p\alpha$)$2n$ with a four-body exit channel, induced by the interaction
of alpha-particles at energy of $E_\alpha=67.2$ MeV, apparently shows the existence of ten resonant states \cite{Povo}. The most part of these states are narrow resonances, as their total width is less than the energy of
a resonance. Fig. (\ref{Expts}) presents the chronological order of experimental data on $^6$He. As it appears from this picture, there is no general consensus on the spectrum and the role of the continuum is 
far from being understood.

On the theoretical side, the 2n-halo structure of $^6$He was investigated in several different formalisms. Many predictions, most of which incomplete in one way or another, suggest a sequence of 
levels $0_{gs}^+, 2_{1}^+, 2_{2}^+, 1^+, 0_{1}^+$, but disagree on the positions and widths of these states.
Most of the $^6$He structure predictions took only $p$-shell excitations into account, but more complete picture must include the promotion of neutrons to $sd-$shell.
In particular $sd-$shell plays a vital role, allowing the possibility of dipole excitations to the continuum.
The halo structure of $^6$He is quite well understood by $^4$He$+n+n$ model. The binding energy is underestimated by a small amount ($\sim0.2$ MeV less than the observed value) and this suggests that 
$^4$He core excitations might be important \cite{Aoy,Cso,Ara}.
In order to understand the weak binding characterstics of light nuclei close to drip line, the continuum coupling effects have been investigated within various frameworks: the Gamow Shell model \cite{Mic1,Mic2,Hage,Mic3},
the Continuum Shell Model \cite{Volya}, the Complex Scaled Cluster Orbital Shell Model \cite{Myo2} and the Hyperspherical Harmonics Expansion \cite{Zhu}. All these nuclear models are successful in predicting the ground state and first excited state structure to a 
reasonable degree, but they disgree for predictions of other higher excited states. 
The Quantum Monte Carlo p-shell calculations \cite{Piep}, along with well established ground state and first excited state structure, predict the energy of the excited $0^+$ state at about $4.66$ MeV, 
depending on the interaction used. In other calculations, the energy of the excited $0^+$ state might be as low as $4.9$ MeV \cite{Myo2} or as high as $8$ MeV \cite{Hage}. 
The energy of the $1^+$ state covers the range of $3.4$ \cite{Hage} to $8$ \cite{Nav1} MeV.
On the other hand, in the few-body calculations of Ref. \cite{Dan1}, the two $0^+$ states were nearly pure jj-coupled states. This calculation allowed excitations into the sd-shell, but these turned out to be 
small for the g.s. and even less for the excited $0^+$ state. The sd-shell occupancy was larger for the $2^+$ states.
In order to avoid the uncertainties due to the the treatment of $\alpha$ particle as point particle, recently Ref. \cite{Mik}
studied the $^6$He nucleus in a fully microscopic six-nucleon calculation, claiming that the E1 strength function exhibits a two-peak structure at around $3$ and $33$ MeV excitation energy. 
The lower peak is well understood in the framework of the  $\alpha+$n$+$n structure and its excitation mechanism is consistent with the classical interpretation of the soft dipole mode (SDM). 
The higher peak is the typical giant dipole resonance that exhibits out-of-phase proton-neutron collective oscillations. Just a few MeV above the SDM peak, some new modes
are found that can be regarded as a vibrational excitation of the SDM.
Most of the theoretical models explains ground state structure fairly well to study dynamics of nuclear reactions, but they lack on incorporation of effects due to the presence of the continuum. 
These effects plays vital role to understand the major reason of their stable character. 
Only very recently in Ref. \cite{Red} the continuum has been included, they found several resonances,including the well-known narrow $2_1^+$ and the recently 
measured broader $2_2^+$. Additional resonant states emerged in the $2^-$and $1^+$ channels near the second $2^+$ resonance and in the $0^-$ channel at
slightly higher energy. 
Motivated by the recent experimental measurements at GANIL \cite{Moug,Povo}, on continuum resonances in $^6$He, we have developed a simple theoretical model \cite{Fort} to study the weakly bound
ground state and low-lying continuum states of $^6$He by coupling two unbound p-waves of $^5$He. In our approach, rather than simulating the resonance with a bound wave function, 
we calculate the full continuum single-particle spectrum of $^5$He in a straightforward fashion and use two copies of the oscillating continuum wave functions to construct two-particle states. In the present study we have extended the model space with 
inclusion of sd- continuum waves of $^5$He. The large basis set of these spd- continuum wavefunctions are used to construct the two-particle $^6$He ground state $0^+$ emerging from five different possible 
configurations i.e. $(s_{1/2})^2$, $(p_{1/2})^2$, $(p_{3/2})^2$, $(d_{3/2})^2$ and $(d_{5/2})^2$. The simple pairing contact-delta interaction is used and pairing strength is adjusted to reproduce the bound ground 
state of $^6$He. The extension of model space is a computationally challenging problem that required careful planning and consideration before undertaking the numerical work. The main aim is to show how an extension of theoretical concepts related to residual interactions, 
namely a contact delta pairing interaction, naturally explain the stable character of the bound states of Borromean nuclei, such as $^6$He and simultaneously account for some of the resonant structures seen in the 
low-lying energy continuum. The extension of model space also allowed us to calculate the monopole, dipole and octupole response of the system.

The paper is organized as follows: 
section $2$ describes the complete formulation of our simple structure model. In section $3$ we analyzed the subsystem $^5$He and section $4$ presents a comparison of our present calculations on ground state properties 
of $^6$He with previous calculations.
Section $5$ describes the procedure that we have adopted for setting the pairing strengths for various multipolarities, followed by mathematical set up for electric transitions to continuum in section $6$.
Section $7-9$ presents the new results on monopole, dipole and octupole response of the system. Finally, section $9$ presents our conclusions. 

\section{Model Formulation}
\label{sec:1}
Each single particle continuum wavefunction of $^5$He is given by
\be
\phi_{\ell,j,m}(\vec{r},E_C)=\phi_{\ell,j}(r,E_C)[Y_{\ell m_\ell}(\Omega)\times \chi_{1/2,m_s}]^{(j)}_m \label{spwfn} 
\ee
The combined tensor product of these two is given by
\be
\psi_{JM}(\vec{r}_1,\vec{r}_2)=[\phi_{\ell_1,j_1,m_1}(\vec r_1,{E_C}_1) \times \phi_{\ell_2,j_2,m_2}(\vec r_2, {E_C}_2)]^{(J)}_M\label{tpwfn}
\ee
In the following we will omit the explicit dependence on the energies ${E_C}_1$ and ${E_C}_2$, although it is understood that each two-particle wavefunction still depends upon two energies. 
In $LS$-coupling for $\ell_1\neq \ell_2$ the antisymmetric wavefunction $\psi\left(\ell_1\ell_2 SLJM\right)$ is given by
\bea
\psi\left(\ell_{1}\ell_{2} SLJM\right)=&&\dfrac{1}{\sqrt{2}}\sum_{M_{S},M_{L}}\langle SM_SLM_L|SLJM\rangle \times
[\phi_{12}(\ell_{1}\ell_{2}LM_L)\chi_{12}(s_{1}s_{2}SM_S)\nonumber\\
&&-\phi_{21}(\ell_{2}\ell_{1}LM_L)\chi_{21}(s_{2}s_{1}SM_S)] \label{awf}
\eea
By making use of symmetry relations Eq.(\ref{awf}) can be written as 
\bea
\psi\left(\ell_{1}\ell_{2} SLJM\right)=&&\dfrac{1}{\sqrt{2}}\sum_{M_S,M_L}\langle SM_SLM_L|SLJM\rangle \times 
[\phi(\ell_{1} \ell_{2}LM_L)+\nonumber \\&&(-1)^{(\ell_{1}+\ell_{2}-L+S)} 
\phi(\ell_{2}\ell_{1}LM_L)]\chi(s_{1}s_{2}SM_S) \label{awf1}
\eea
The generic matrix elements (diagonal and non-diagonal) due to mutual interaction $V_{12}$ in $LS$-coupling of two particles are given by
\bea
&&\langle\ell_a\ell_bSLJM|V_{12}|\ell_c\ell_dS'L'J'M'\rangle=
\sum\langle SM_SLM_L|SLJM\rangle\langle S'M'_SL'M'_L|S'L'J'M'\rangle \nonumber \\
&&\langle s_1m_{s_1}s_2m_{s_2}|s_1s_2SM_S\rangle\langle s'_1m'_{s_1}s'_2m'_{s_2}|s'_1s'_2S'M'_S\rangle  
\langle \ell_am_a\ell_bm_b|\ell_a\ell_bLM_L\rangle \langle \ell_cm_c\ell_dm_d|\ell_c\ell_dL'M'_L\rangle \nonumber \\
&&\int[\phi_1(a)\chi_1(m_{s_1})\phi_2(b)\chi_2(m_{s_2})]^*V_{12}
[\phi_1(c)\chi_1(m'_{s_1})\phi_2(d)\chi_2(m'_{s_2})]d\vec{r}_1d\vec{r}_2 \label{me5}
\eea
where quantum numbers $\ell_a$ and $\ell_c$ are associated with particle $1$, $\ell_b$ and $\ell_d$ 
are associated with particle $2$. $\ell_a$ and $\ell_b$ are coupled to L and $\ell_c$ and $\ell_d$ are 
coupled to $L'$. 
We take an attractive pairing contact delta interaction because 
we can reach the goal of calculation of electromagnetic response with only a few parameters
(the pairing strengths). For $S=0$ the explicit expression for $V_{12}$ is given by
\be
\delta\left(\vec{r_1}-\vec{r_2}\right)=\frac{\delta\left(r_1-r_2\right)}{r_1r_2}\delta\left(\cos{(\theta_1)}-\cos{(\theta_2)}\right)\delta\left(\varphi_1-\varphi_2\right)
\label{me7} 
\ee
Using Eq.(\ref{spwfn}) and Eq.(\ref{me7}) and making use of the fact that $V_{12}$ is spin independent,
the integral in Eq.(\ref{me5}) can be rewritten as 
\bea
&&\int[\phi_1(a)\chi_1(m_{s_1})\phi_2(b)\chi_2(m_{s_2})]^*V_{12}
*[\phi_1(c)\chi_1(m'_{s_1})\phi_2(d)\chi_2(m'_{s_2})]d\vec{r}_1d\vec{r}_2 =\nonumber\\
&&\int R_{n_a\ell_a}^*\left(r\right)R_{n_b\ell_b}^*\left(r\right)\frac{1}{r^2}R_{n_c\ell_c}\left(r\right)R_{n_d\ell_d}\left(r\right) dr 
\int Y_{\ell_am_a}^*\left(\Omega\right)Y_{\ell_bm_b}^*\left(\Omega\right)Y_{\ell_cm_c}\left(\Omega\right)Y_{\ell_dm_d}\left(\Omega\right)d\Omega \label{me8}
\eea
Using the property of two spherical harmonics of same angles, we have
\bea
&&\left[Y_{\ell_am_a}\left(\Omega\right)Y_{\ell_bm_b}\left(\Omega\right)\right]^*=\sum_{\ell,m}\left(-1\right)^{\ell-m}
\begin{pmatrix}
  \ell & \ell_a & \ell_b \\
  -m & m_a & m_b 
\end{pmatrix} 
\langle \ell\|Y_{\ell_a}\|\ell_b\rangle^* Y_{\ell m}\left(\Omega\right)^* \nonumber \\
&&\left[Y_{\ell_cm_c}\left(\Omega\right)Y_{\ell_dm_d}\left(\Omega\right)\right] =\sum_{\ell',m'}\left(-1\right)^{\ell'-m'}
\begin{pmatrix}
  \ell' & \ell_c & \ell_d \\
  -m' & m_c & m_d 
\end{pmatrix}
\langle \ell'\|Y_{\ell_c}\|\ell_d\rangle Y_{\ell'm'}\left(\Omega\right) \label{me9}
\eea
where the Condon and Shortley phase convention has been adopted.
Using the orthonormality property of spherical harmonics i.e.
\begin{equation}
\int  Y_{\ell m}^* Y_{\ell'm'}d\Omega=\delta_{\ell \ell'}\delta_{mm'} \label{me10}
\end{equation}
we are left with
\bea
\int Y_{\ell_am_a}^*\left(\Omega\right)Y_{\ell_bm_b}^*\left(\Omega\right)Y_{\ell_cm_c}\left(\Omega\right)Y_{\ell_dm_d}\left(\Omega\right)d\Omega=&&  
\sum_{\ell,m}(-1)^{2(\ell-m)}
\begin{pmatrix}
  \ell & \ell_a & \ell_b \\
  -m & m_a & m_b 
\end{pmatrix}
\begin{pmatrix}
  \ell & \ell_c & \ell_d \\
  -m & m_c & m_d 
\end{pmatrix}\nonumber \\ 
&&\langle \ell\|Y_{\ell_a}\|\ell_b\rangle^*\langle \ell'\|Y_{\ell_c}\|\ell_d\rangle \label{me11}
\eea
Hence, using the above assumptions and properties, Eq.(\ref{me5}) is reduced to
\begin{eqnarray}
&&\langle\ell_a\ell_bSLJM|V_{12}|\ell_c\ell_dS'L'J'M'\rangle=
\sum\langle SM_SLM_L|SLJM\rangle\langle S'M'_SL'M'_L|S'L'J'M'\rangle \nonumber \\ 
&&\langle s_1m_{s_1}s_2m_{s_2}|s_1s_2SM_S\rangle\langle s'_1m'_{s_1}s'_2m'_{s_2}|s'_1s'_2S'M'_S\rangle  
\langle \ell_am_a\ell_bm_b|\ell_a\ell_bLM_L\rangle \langle \ell_cm_c\ell_dm_d|\ell_c\ell_dL'M'_L\rangle \nonumber \\
&&\sum_{\ell m}(-1)^{2(\ell-m)}
\begin{pmatrix}
  \ell & \ell_a & \ell_b \\
  -m & m_a & m_b 
\end{pmatrix}
\begin{pmatrix}
  \ell & \ell_c & \ell_d \\
  -m & m_c & m_d 
\end{pmatrix}\nonumber\\
&&\langle \ell\|Y_{\ell_a}\|\ell_b\rangle^*\langle \ell'\|Y_{\ell_c}\|\ell_d\rangle 
\int R_{n_a\ell_a}^*\left(r\right)R_{n_b\ell_b}^*\left(r\right)\frac{1}{r^2}R_{n_c\ell_c}\left(r\right)R_{n_d\ell_d}\left(r\right) dr \label{me12}
\end{eqnarray}
The major ingredients for the complete study of $^6$He are the matrix elements of pairing interaction.
These correspond to the radial integrals and to the coefficients.
The coefficients of these matrix elements of eq. (\ref{me12}) for $0^+, 1^-, 2^+$ and $3^-$ are summarized in tables \cite{JS2}.
The full computational procedure is described in details in \cite{JS2, JST}.

\section{Analysis of $^4$He$+n$ subsystem }
Analysis of the $^4$He$+n$ subsystem ($^5$He) is indespensable in studying $^6$He as a typical nucleus of Borromean system of $^4$He$+n+n$.
In order to study the the binding mechanism of $^6$He, the interaction between a core of $^4$He and a valence neutron palys an important role.
The unbound nucleus $^5$He can be described as an inert $^4$He core with an unbound neutron moving in $p$, $d$ or $s$ states in simple independent-particle shell model picture.
These $p$ and $d$ states are split by spin-orbit interaction. Experimentally only the $p_{3/2}$ and $p_{1/2}$ resonances are confirmed at $0.789$ and $1.27$ MeV respectively above the neutron sepration threshold.
Their widths are quoted as $0.648$ MeV and $5.57$ MeV respectively \cite{TUNL}.
Theoretically in order to extend the model space we have also included the $sd-$shell in picture.
The continuum monopole ($\ell=0$), dipole ($\ell=1$) and quadrupole ($\ell=2$) scattering single particle states ($E_C>0, k> 0$) of $^5$He are generated with Woods-Saxon (WS) potential given by
\bea
 V_{WS}=\left[ V_0 + V_{ls}r_0^2(\vv{l}.\vv{s})\frac{1}{r}\frac{d}{dr}\right]\left[1+exp\left(\frac{r_0-R}{a}\right)\right]^{-1} \label{WSP}
 \eea
where $R=r_0A^\frac{1}{3}$. For $^5$He the parameter set used is WS potential depth $V_0=-42.6$ MeV, $r_0=1.2$ fm, $a=0.9$ fm and spin-orbit coefficient $V_{ls}=8.5$ MeV. 
The continuum single-particle wavefunctions are calculated (see Fig. $2$ of \cite{Fort}, Fig. $1$ of \cite{JS1} and Fig. $2$ of \cite{JS2}), with energies from 0.0 to 10.0 MeV, 
normalized to a Dirac delta in energy, for the spd-states of $^5$He on a radial grid that goes from 0.1 fm to 100.0 fm with the 
potential given above.

\section{Ground state properties of $^6$He}
The simple model with two non-interacting particles in the above single-particle levels of $^5$He produces
different parity states (see Table-$1$ of \cite{JS1}) when two neutrons are placed in five different unbound orbits, $s_{1/2}$, $p_{1/2}$, $p_{3/2}$, 
$d_{3/2}$ and $d_{5/2}$. Namely five configurations $(s_{1/2})^2$, $(p_{1/2})^2$, $(p_{3/2})^2$, $(d_{3/2})^2$ 
and $(d_{5/2})^2$ couple to $J=0^+$, seven configurations $(s_{1/2}d_{3/2})$, $(s_{1/2}d_{5/2})$, $(p_{1/2}p_{3/2})$, $(p_{3/2}p_{3/2})$, $(d_{3/2}d_{3/2})$, 
$(d_{3/2}d_{5/2})$ and $(d_{5/2}d_{5/2})$ couple to $J=2^+$, five configurations $(s_{1/2}p_{1/2})$, $(s_{1/2}p_{3/2})$, $(p_{1/2}d_{3/2})$, $(p_{3/2}d_{3/2})$ and 
$(p_{3/2}d_{5/2})$ couple to $J=1^-$ and three configurations $(p_{1/2}d_{5/2})$, $(p_{3/2}d_{3/2})$ and $(p_{3/2}d_{5/2})$ couple to $J=3^-$. Other less important multipolarities can also be constructed as in 
Table-$1$ of \cite{JS1}.

An attractive pairing contact delta interaction has been used, $-g\delta(\vec r_1 - \vec r_2)$ for simplicity, because we 
can reach the goal with only one parameter adjustment. 
 With continuum single-particle wavefunctions, using the mid-point method with an energy spacing of 
2.0, 1.0, 0.5, 0.2 and 0.1 MeV, corresponding to block basis dimensions of $N=$5, 10, 20, 50 and 100 respectively, 
the two particle states are formed and the matrix elements of the pairing interaction are calculated. 
The resulting matrix has been diagonalized with standard routines and it has given the eigenvalues shown 
in Fig. $2$ of \cite{JS1} for the $J=0^+$ case. It is clear from eigenspectrum that, with increase in basis 
dimensions the superflous bound states moves into the continuum. This was not present in our old work \cite{Fort}, and it can be attributed to the new, more complete basis.
The coefficient of the $\delta-$contact matrix, $G$, has been adjusted to reproduce the correct ground state energy each time. 
The actual pairing interaction $g$ is obtained by correcting with a factor that depends on the 
aforementioned spacing between energy states and it is practically a constant, 
except for the smallest basis.  The biggest adopted basis size gives a fairly dense continuum in 
the region of interest.
The radial part of the $S=0$ g.s. wavefunction (see Fig. $3$ of \cite{JS1}) obtained from the diagonalization in the largest basis, shows a certain degree of collectivity, taking contributions of comparable magnitude from several basis states, 
while in contrast the remaining unbound states usually are made up of a few major components \cite{JST}.
The detailed components for each configuration is summarized in Table-\ref{comp}, and compared with the 
previous calculations of Myo \cite{MYO} and Hagino \cite{HAG}. Present calculations are well in agreement with
previous calculations.
The calculated ground state properties are summarized in Table-\ref{gsp} in comparison with calculations \cite{MYO,HAG}, where $R_{m}$ is the matter radius,
\begin{equation}
\langle r_{NN}^2\rangle = \langle \psi_{gs}(\vec{r}_1,\vec{r}_2) |(\vec{r}_1-\vec{r}_2)^2| \psi_{gs}(\vec{r}_1,\vec{r}_2) \rangle
\end{equation}
is the mean square distance between the valence neutrons, and
\begin{equation}
\langle r_{c-NN}^2\rangle = \langle \psi_{gs}(\vec{r}_1,\vec{r}_2) |(\vec{r}_1+\vec{r}_2)^2/4| \psi_{gs}(\vec{r}_1,\vec{r}_2) \rangle
\end{equation}
is the mean square distance of their centre of mass with respect to the core.
\begin{table}[h]
\centering
\caption{Components of the ground state ($0^+_1$) of $^{6}$He}
\centering
\vspace*{0.3cm}
\begin{tabular}{cccc}
\hline
\vspace*{0.3cm}
Config.         &  Present & T.Myo\cite{MYO} & Hagino\cite{HAG} \\ 
\hline
 $(2s_{1/2})^2$  &  0.008           & 0.009 & -- \\
 $(1p_{1/2})^2$  &  0.080          & 0.043  & -- \\
 $(1p_{3/2})^2$  &  0.897           & 0.917 & 0.830\\
 $(1d_{3/2})^2$  &  0.005           & 0.007 &  --\\
 $(1d_{5/2})^2$  &  0.009           & 0.024 &  --\\
\hline
\label{comp}
\end{tabular}
\end{table}
\begin{table}[h]
\centering
\caption{Radial properties of the ground state of $^{6}$He in units of fm}
\centering
\vspace*{0.3cm}
\begin{tabular}{cccc}
\hline
\vspace*{0.3cm}
         &  Present & T.Myo\cite{MYO} & Hagino\cite{HAG} \\ 
\hline
$R_{m}$   &  $2.37674$           & $2.37$ & ... \\   
 $r_{NN}^2$  &  $28.8404$           & $23.2324$ & $21.3$ \\
 $r_{c-2N}^2$  &  $7.21011$          & $9.9225$  & $13.2$ \\
\hline
\label{gsp}
\end{tabular}
\end{table}
In Table-\ref{gsp}, while the matter radius is consistent with that of Myo, there are large differences for the other two quantities that can be ascribed to the choice of different pairing interactions.
The two particle density of $^6$He as a function of two radial coordinates, $r_1$ and $r_2$, for valence neutrons, and the angle between them, $\theta_{12}$ in LS-coupling scheme is given by
\begin{figure}[!t]
\begin{center}
\vspace*{-1.3cm}
\hspace*{-2.4cm}
\includegraphics[width=0.75\textwidth, clip=]{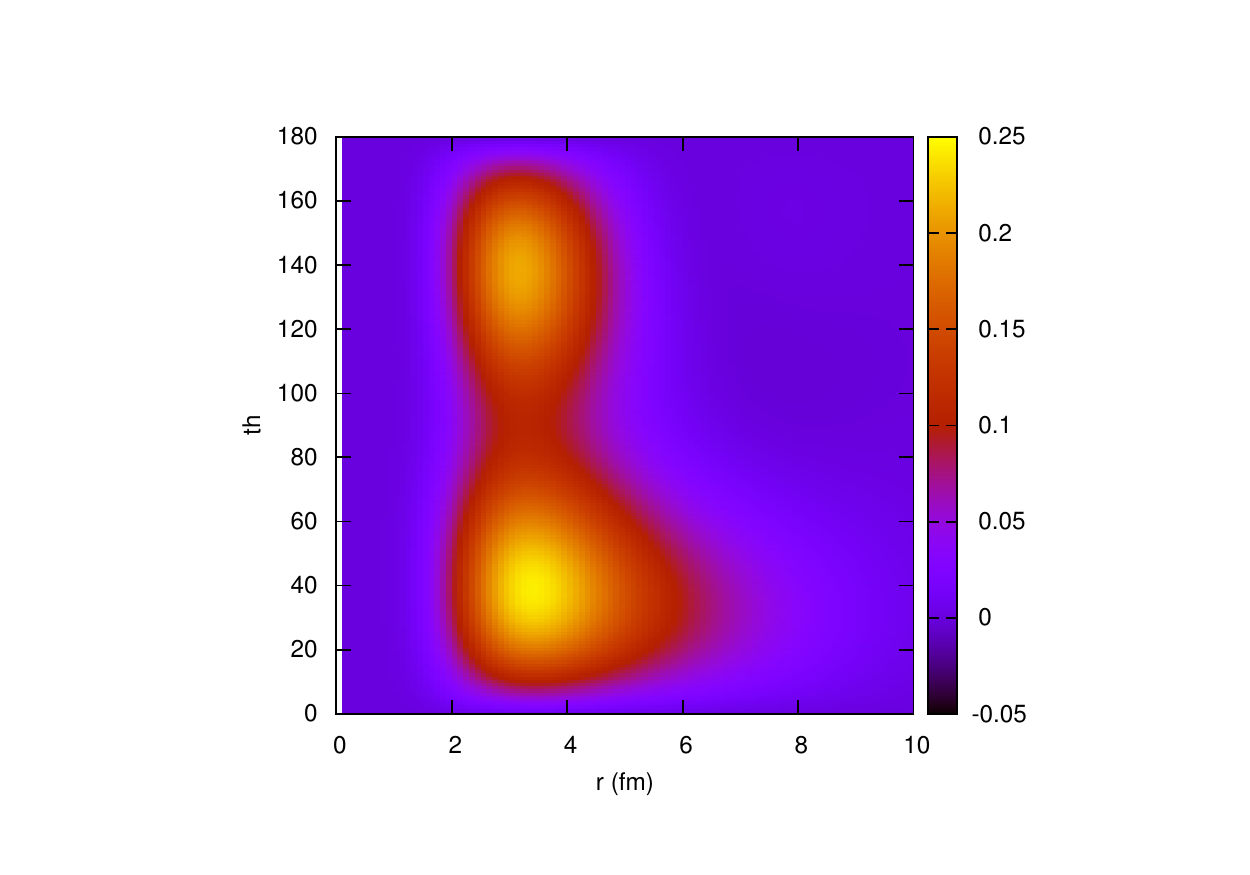} 
\end{center}
\vspace*{-1.4cm}
\caption{(Color online) Two-particle density for $^6$He as a function of $r_1=r_2=r$ and angle between the valence neutrons $\theta_{12}$.}
\label{Den}
\end{figure}
\be
\rho(r_1, r_2, \theta_{12})=\rho^{S=0}(r_1, r_2, \theta_{12})+\rho^{S=1}(r_1, r_2, \theta_{12})
\ee
The explicit expression for $S=0$ component is given by \cite{Ber}
\bea
\rho^{S=0}(r_1, r_2, \theta_{12})=  &&\frac{1}{8\pi}\sum_{L}\sum_{\ell,j}\sum_{\ell',j'}\frac{\hat{\ell}\hat{\ell'}\hat{L}}{\sqrt{4\pi}} 
\begin{pmatrix}
  \ell & \ell' & L \\
  0 & 0 & 0 
\end{pmatrix}^2  
\times \psi_{\ell j}(r_1,r_2) \psi_{\ell'j'}(r_1,r_2) Y_{L0}(\theta_{12})\nonumber \\ 
&& \times (-1)^{\ell+\ell'}\sqrt{\frac{2j+1}{2\ell+1}}\sqrt{\frac{2j'+1}{2\ell'+1}}
\eea
where $\hat{\ell}=\sqrt{2l+1}$ and $ \psi_{\ell j}(r_1,r_2)$ is the radial part of two particle wavefunction given by
\bea
 \psi_{\ell j}(r_1,r_2)= \sum_{n2\leq n1}\frac{\alpha_{n_1n_2\ell j}}{\sqrt{2(1+\delta_{n_1n_2})}} 
 \times (\phi_{n_1\ell j}(r_1)\phi_{n_2\ell j}(r_2)+\phi_{n_1\ell j}(r_2)\phi_{n_2\ell j}(r_1))
\eea
where $n_1$ and $n_2$ are radial quantum numbers and $\alpha_{n_1n_2\ell j}$ is an expansion coefficient. 
Fig. (\ref{Den}) shows the two-particle density plotted as a function of the radius $r_1=r_2\equiv r$ and the angle $\theta_{12}$, and with a
weight factor of $4\pi r^2\cdot2\pi r^2$sin$\theta_{12}$. As it has been pointed out in \cite{HAG}, one observes two peaks in the two particle densities.
The peak at smaller and larger $\theta_{12}$ are referred to as \textquotedblleft di-neutron\textquotedblright and \textquotedblleft cigar-like\textquotedblright 
configurations respectively. In this case the di-neutron component has a slightly higher density and it has a longer radial tail, which confirms the halo structure of $^6$He, while the cigar-like component has a very compact structure
comparatively. The percentage contribution of di-neutron configuration is $\sim 64\%$, while the cigar component has $\sim 36\%$ contribution.

\section{Pairing strength of different multipolarities}
\begin{figure}[!h]
\begin{center}
\includegraphics[width=0.7\textwidth]{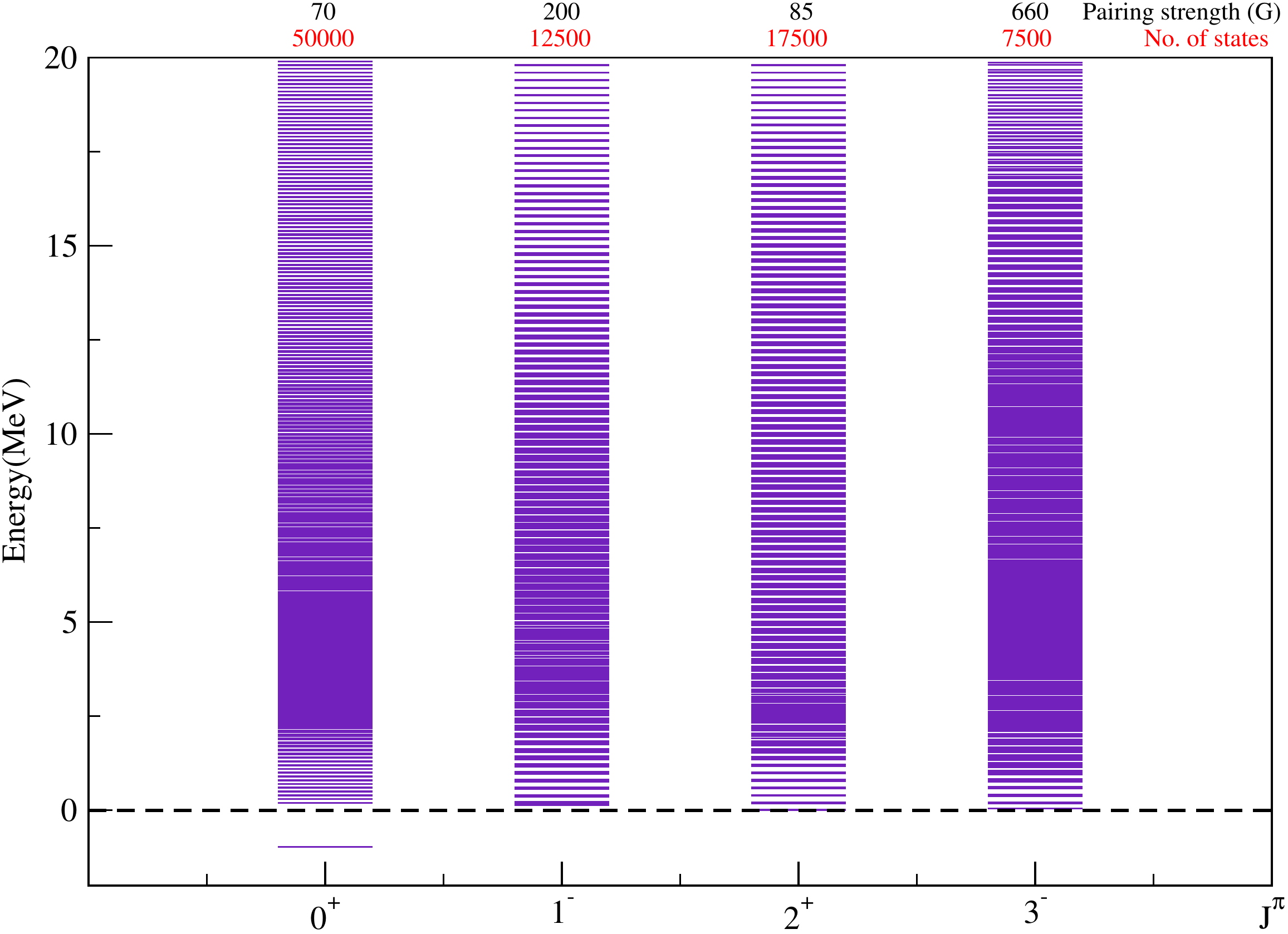}
\end{center}
\caption{(Color online) Eigenspectrum of the interacting two-particle case for J$^\pi=0^+$, $1^-$, $2^+$ and $3^-$  for different number of states. 
The coefficient of the $\delta-$contact matrix, $G$, has also been shown for different J.}
\label{ES}
\end{figure}

Theoretical investigation of very weakly-bound nuclei sitting right on top of the drip lines, demands proper consideration of nucleon-nucleon pairing interaction.
An attractive pairing contact delta interaction has been used, $-G\delta(\vec r_1 - \vec r_2)$ for simplicity, because we can reach the
goal with only one parameter adjustment. 
For ground state it is pretty much clear that, the pairing strength, G, is adjusted in order to get the correct ground state energy. But for higher multipolarities
i.e. J$=1^-$, $2^+$ and $3^-$ we do not have a clear-cut strategy to determine the exact value of pairing strength. This is the main reason why we did not enter into the 
complications of a density dependent pairing interaction: there is no unique way to adjust the parameters and geometry. 
For each value of J we tried different sets of values of G. From Fig-\ref{ES}, the upper limit of pairing strength can be 
found for several values of J, along with the number of states (red). 
Notice that different multipolarities give rise to different concentrations of strength as seen by comparing the densities of the various columns.  
Notice also that the continua are, at the eyes, quite dense, a condition that is necessary to reproduce minute features with the necessary accuracy.
\section[Electric trans..]{Electric transitions to continuum- Mathematical set up}
The electric transition probability amplitude between ground state $\psi(j_{1}^{'},j_{2}^{'},J^{'},M^{'})$ and continuum states $\psi(j_{1},j_{2},J,M)$ is given by
\bea
&& \langle\psi(j_{1}^{'},j_{2}^{'},J^{'},M^{'})|\hat{O}_p|\psi(j_{1},j_{2},J,M)\rangle = 
\sum_{S^{'},L^{'}}\sqrt{(2S^{'}+1)(2L^{'}+1)(2j_{1}^{'}+1)(2j_{2}^{'}+1)}\nonumber\\
&&\begin{Bmatrix}
1/2 & \ell_{1}^{'} & j_1^{'} \\
1/2 & \ell_{2}^{'} & j_2^{'} \\
S^{'} & L^{'} & J^{'}
\end{Bmatrix}
\sum_{S,L} \sqrt{(2S+1)(2L+1)(2j_{1}+1)(2j_{2}+1)}
\begin{Bmatrix}
1/2 & \ell_{1} & j_1 \\
1/2 & \ell_{2} & j_2 \\
S & L & J
\end{Bmatrix}\nonumber\\ 
&&\bigg([\langle R_{\ell_{1}^{'}\ell_{2}^{'}}^{+}(r_1r_2)\varUpsilon_{L^{'}M^{'}}^{+}(\Omega_1\Omega_2)|\hat{O}_p|
R_{\ell_{1}\ell_{2}}^{+}(r_1r_2)\varUpsilon_{LM}^{+}(\Omega_1\Omega_2)\rangle]+\nonumber\\
&&[\langle R_{\ell_{1}^{'}\ell_{2}^{'}}^{-}(r_1r_2)\varUpsilon_{L^{'}M^{'}}^{-}(\Omega_1\Omega_2)|\hat{O}_p|
R_{\ell_{1}\ell_{2}}^{-}(r_1r_2)\varUpsilon_{LM}^{-}(\Omega_1\Omega_2)\rangle]\bigg)
\label{em1}
\eea
where 
$\hat{O}_p$ is a generic one body operator given by 
\be
\hat{O}_p=e_{eff}^{(\lambda)}\bigg(r_1^{\lambda}Y_{\lambda\mu}(\hat{r}_1)+r_2^{\lambda}Y_{\lambda\mu}(\hat{r}_2)\bigg)\label{em2}
\ee
with $\lambda=1$ for dipole, $\lambda=2$ for quadrupole and $\lambda=3$ for octupole, $e_{eff}^{(\lambda)}$ is the the effective charge, tabulated in Table- (\ref{EC}) for different multipolarities and is given by
\be
e_{eff}^{(\lambda)}=\frac{A_1^{\lambda}Z_2+(-1)^{\lambda}A_2^{\lambda}Z_1}{A^{\lambda}}\label{em22}
\ee
where we use the masses and charges of the $\alpha$-particle and of a neutron for $1, 2$ because the one body operator acts only on one particle at any one time
\begin{table}[!h]
\centering
\caption{Effective charge for different multipolarities.}
\label{EC}
\begin{tabular}{cc}
\hline
$\lambda$        & $(e_{eff}^{(\lambda)})^2$ \\ \hline
$0$ (Monopole)   & $4$                   \\
$1$ (Dipole)     & $4/25$                \\
$2$ (Quadrupole) & $4/625$               \\
$3$ (Octupole)   & $4/15625$             \\ \hline
\end{tabular}
\end{table}
Using Eq.(\ref{em2}), Eq.(\ref{em1}) can be rewritten as
\bea
&& \langle\psi(j_{1}^{'},j_{2}^{'},J^{'},M^{'})|\hat{O}_p|\psi(j_{1},j_{2},J,M)\rangle=
\sum_{S^{'},L^{'}}\sqrt{(2S^{'}+1)(2L^{'}+1)(2j_{1}^{'}+1)(2j_{2}^{'}+1)}\nonumber\\
&&\begin{Bmatrix}
1/2 & \ell_{1}^{'} & j_1^{'} \\
1/2 & \ell_{2}^{'} & j_2^{'} \\
S^{'} & L^{'} & J^{'}
\end{Bmatrix}
\sum_{S,L} \sqrt{(2S+1)(2L+1)(2j_{1}+1)(2j_{2}+1)} 
\begin{Bmatrix}
1/2 & \ell_{1} & j_1 \\
1/2 & \ell_{2} & j_2 \\
S & L & J
\end{Bmatrix} \nonumber\\
&&2 \bigg(\iint R_{\ell_{1}^{'}\ell_{2}^{'}}^{+}(r_1r_2)r_{1}^{\lambda}R_{\ell_{1}\ell_{2}}^{+}(r_1r_2)r_{1}^{2}dr_{1}r_{2}^{2}dr_{2}
\langle \varUpsilon_{L^{'}M^{'}}^{+}(\Omega_1\Omega_2)|Y_{\lambda\mu}(\Omega_1)|\varUpsilon_{LM}^{+}(\Omega_1\Omega_2)\rangle\nonumber\\
 &&+\iint R_{\ell_{1}^{'}\ell_{2}^{'}}^{-}(r_1r_2)r_{1}^{\lambda}R_{\ell_{1}\ell_{2}}^{-}(r_1r_2)r_{1}^{2}dr_{1}r_{2}^{2}dr_{2}
\langle\varUpsilon_{L^{'}M^{'}}^{-}(\Omega_1\Omega_2)|Y_{\lambda\mu}(\Omega_1)|\varUpsilon_{LM}^{-}(\Omega_1\Omega_2)\rangle\bigg)\label{em3}
\eea
Also $R_{\ell_{1}^{'}\ell_{2}^{'}}^{\pm}(r_1r_2)$ and $\varUpsilon_{L^{'}M^{'}}^{\pm}$ are given by
\bea
R_{\ell_{1}\ell_{2}}^{\pm}(r_1r_2)=\frac{1}{r_1r_2\sqrt{2}}
\left[R_{n_{1}\ell_{1}}(r_1)R_{n_{2}\ell_{2}}(r_2)\pm R_{n_{2}\ell_{2}}(r_1)R_{n_{1}\ell_{1}}(r_2)\right]\label{em4}
\eea
\bea
\varUpsilon_{LM}^{\pm}= \frac{1}{\sqrt{2}}\sum\langle\ell_{1}m_1\ell_{2}m_2|\ell_{1}\ell_{2}LM\rangle
\left[Y_{\ell_{1}m_1}(\Omega_1)Y{\ell_{2}m_2}(\Omega_2)\pm Y_{\ell_{2}m_2}(\Omega_1)Y_{\ell_{1}m_1}(\Omega_2)\right]\label{em5}
\eea
Using Eq.(\ref{em4}) and Eq. (\ref{em5}), Eq.(\ref{em3}) gives us the matrix elements of different multipolarities.
Eq.(\ref{em3}) consists of two parts i.e. evaluation of radial parts and angular parts.
For evaluation of radial integrals, we need the corresponding two-particle wave function, whereas for the angular part
by making use of Eq.(\ref{em5}), we will simplify the angular part and for different multipolarities these can be easily calculated.

Clearly our calculations give strength distributions at discrete values of energy to which we apply a Gaussian smoothing procedure that does not alter the total integrated strength \footnote{There is a minor loss
of strength close to zero that could be avoided by using for example Lorentzian functions instead of Gaussians.}.

\section{Monopole strength distribution}
\begin{figure}[!t]
\begin{center}
\includegraphics[width=0.60\textwidth]{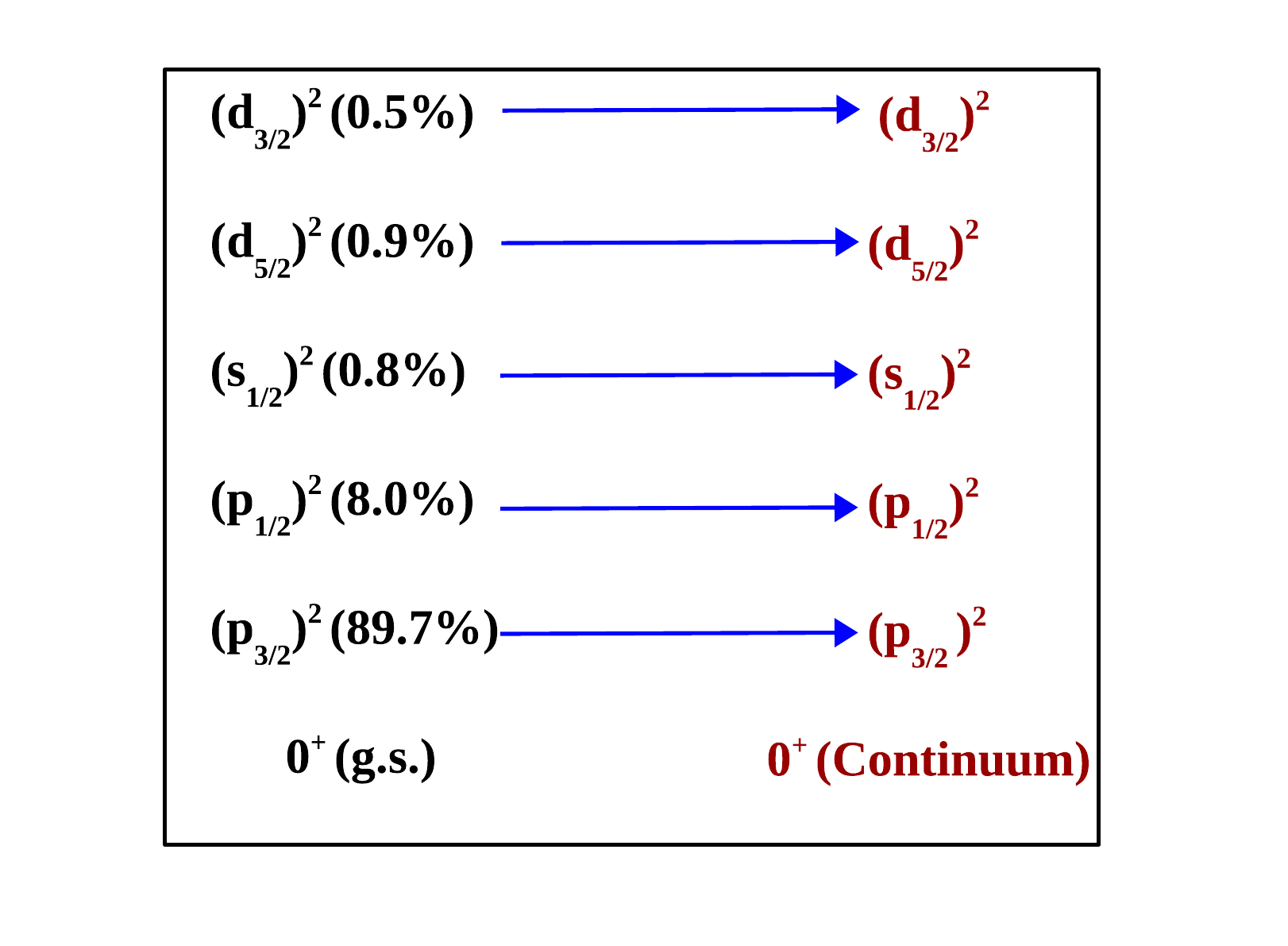}
\end{center}
\caption{(Color online) Total number of possible monopole transitions from ground state $0^+$ to the final continuum $0^+$ states with different contributions from five different possible configurations for $^6$He.}
\label{Mon2}
\end{figure}

\begin{figure}[!h]
\begin{center}
\includegraphics[width=0.70\textwidth]{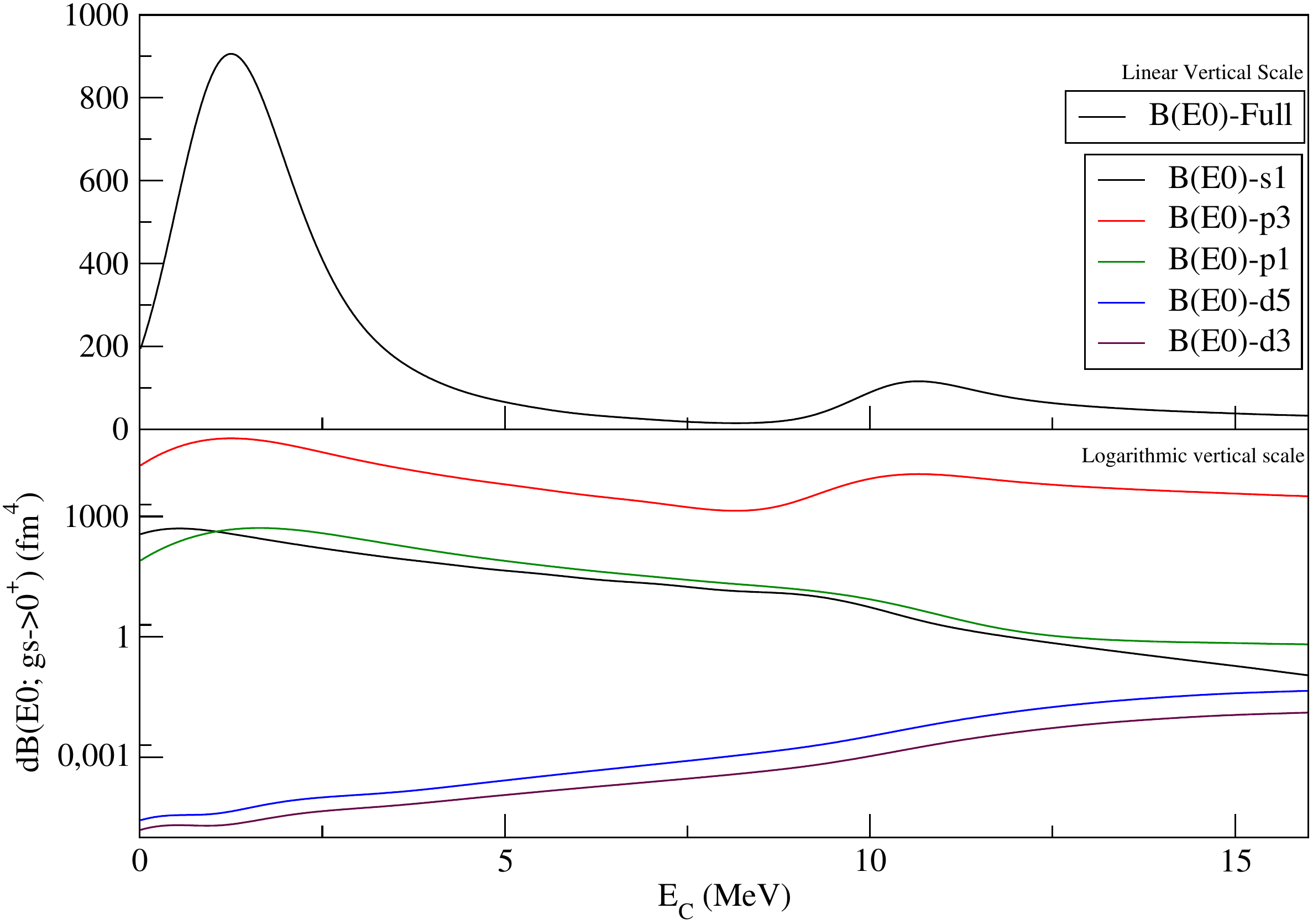}
\end{center}
\caption{(Color online) (Upper panel) Total monopole $E0$ transition strength distribution (on linear vertical scale) from ground state $0^+$ to the final state $0^+$ for $^6$He.
(Lower panel) Component monopole $E0$ transition strength distribution (on logarithmic vertical scale) from ground state $0^+$ to the final state $0^+$ for $^6$He. }
\label{Mon3}
\end{figure}

Electric monopole transition strengths reflect the off diagonal matrix elements of the
E0 operator. The E0 operator \cite{Kant} can be expressed in terms of single-nucleon degrees of freedom as
\be
\hat{T}(E0)=\sum_{k}e_kr_k^2
\ee
The E0 transition rate, $1/\tau(E0)=\rho_{fi}^2$, is defined by 
\be
\rho_{fi}^2=\left|\frac{\langle f|\sum_{k}e_kr_k^2|i\rangle}{eR^2}\right|^2
\ee
where, $e$ is the unit of electrical charge, and R is the nuclear radius, $R\backsimeq1.2A^{1/3}$ fm. 
These calculations also leads us to study the role of various configurations in the total monopole strength.
After constructing a basis of the largest size (N$=100$) made up of five parts, namely $[s_{1/2}^2]^{(0)}$, $[p_{1/2}^2]^{(0)}$, $[p_{3/2}^2]^{(0)}$, $[d_{3/2}^2]^{(0)}$ and $[d_{5/2}^2]^{(0)}$, we diagonalize the pairing matrix and obtain eigenvalues and eigenvectors for $J=0$. Only one state is bound and all the remaining ones are unbound. 
In order to reduce computation time, we have performed a set of calculations for monopole transitions from ground 
state $0^+$ for basis size N$=100$ to the continuum $0^+$ for basis size N$=50$. From Fig. (\ref{Mon2}), it is clear that there are only
five possible transitions from $0^+$ ground state components to continuum $0^+$ states components.
With all these necessary ingredients i.e. ground state and continuum $0^+$ states, the monopole strength distribution has been studied. The upper panel of 
Fig. (\ref{Mon3}), shows the total monopole transition strength of $^6$He and lower panel of Fig. (\ref{Mon3}), shows the contribution of various possible transitions on logarithmic scale.
From lower panel Fig. (\ref{Mon3}), it is clear that the transition $[(p_{3/2})^2]^{(0)}$(g.s.)$\rightarrow$ $[(p_{3/2})^2]^{(0)}$(continuum), is dominant in the monopole transition strength, whereas 
the transition $[(d_{3/2})^2]^{(0)}$(g.s.)$\rightarrow$ $[(d_{3/2})^2]^{(0)}$(continuum) is the least significant in total monopole transition strength.
From this, one can also see that transition $[(s_{1/2})^2]^{(0)}$ (g.s.)$ \rightarrow $ $[(s_{1/2})^2]^{(0)}$ (continuum) has significant contribution to the total strength, which justifies the 
inclusion of sd- shell in calculations. 
The total integrated monopole strength amounts to about $2682.97 fm^4$. 
This value can be compared with the non energy weighted sum rule calculations for monopole strength, that gives about $2800 fm^4$, using formulas of Ref. \cite{Mey}, giving a very good agreement.


\section{Dipole strength distribution}
\begin{figure}[!h]
\begin{center}
\includegraphics[width=0.60\textwidth]{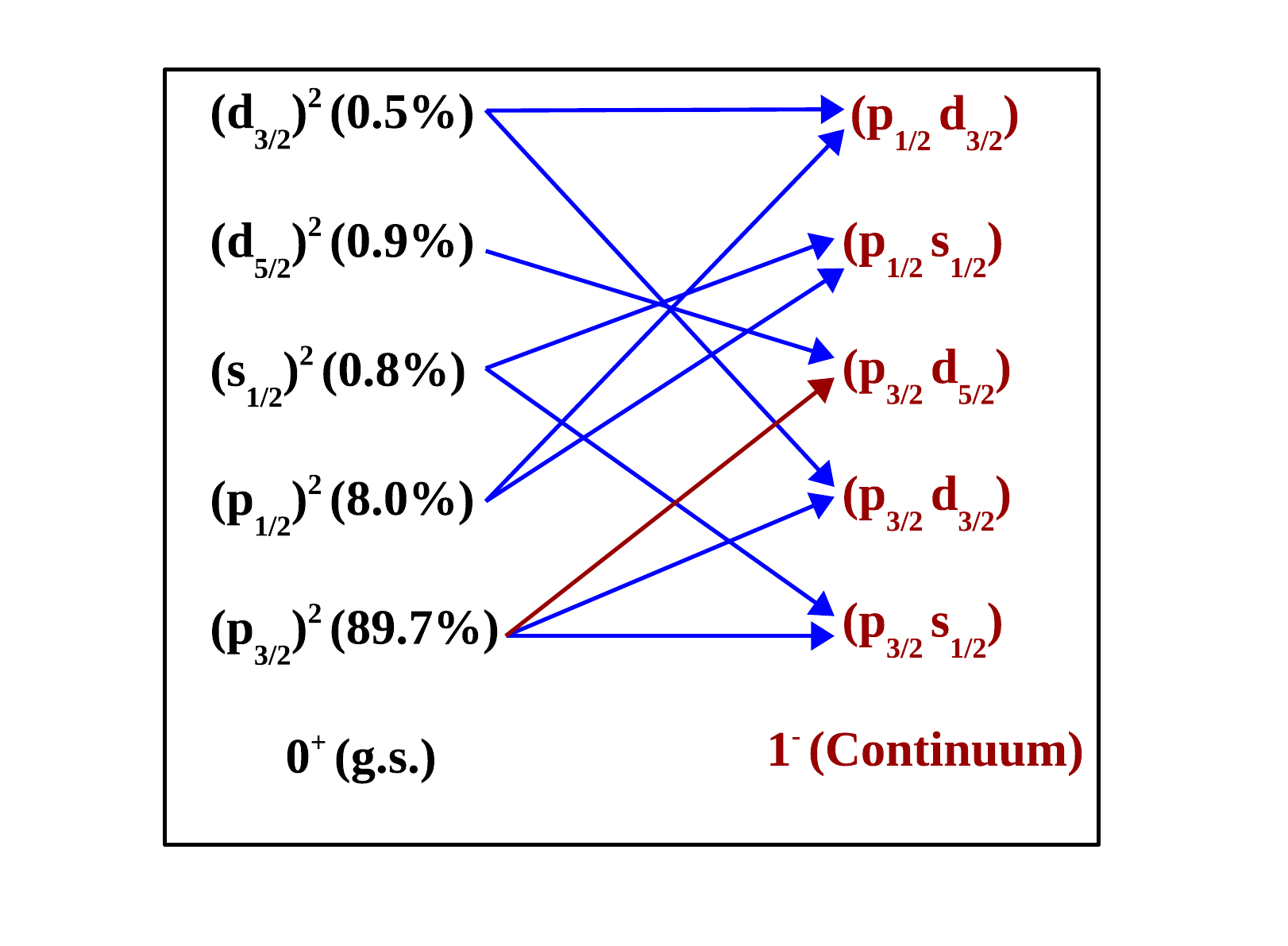}
\end{center}
\caption{(Color online) Total number of possible dipole transitions from ground state $0^+$ to the final state $1^-$ with different contributions from five different possible configurations for $^6$He. 
The dominant transition is highlighted in red color.}
\label{Dip1}
\end{figure}

\begin{figure}[!h]
\begin{center}
\includegraphics[width=0.70\textwidth]{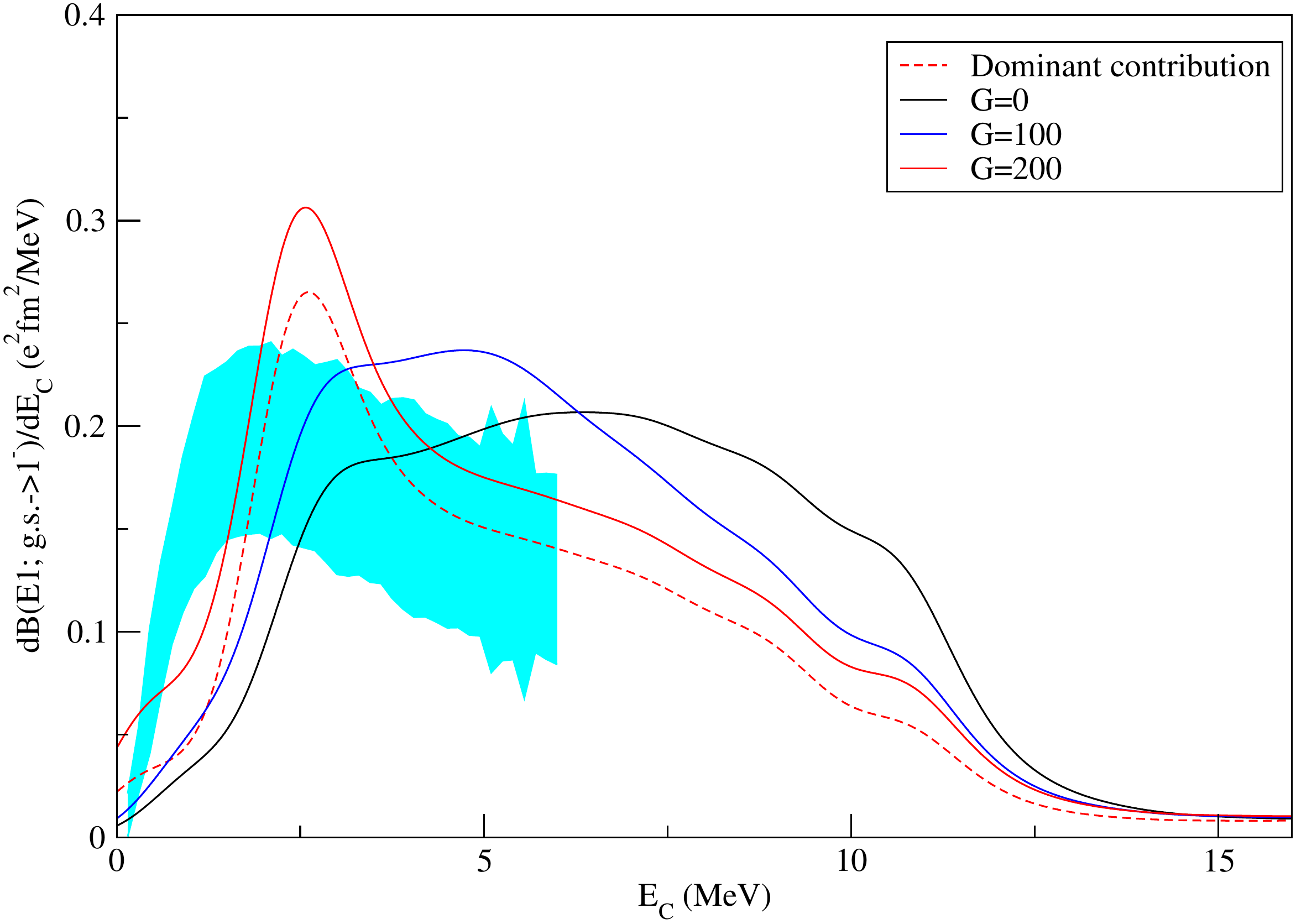}
\end{center}
\caption{(Color online) Dipole $E1$ transition strength distribution from ground state $0^+$ to the final state $1^-$ for $^6$He for a few values of the pairing strength compared to experimental data up to $6$ MeV from ref \cite{Aum}.}
\label{Dip2}
\end{figure}

While most theoretical studies have been focused on dipole strength \cite{Mik, Aoy3, Desco}, our includes many more multipolarities. 
In order to compare our approach with others, we have also performed 
a set of calculations for dipole response from ground state to all components of $1^-$ state. 
After constructing a basis of the dimensions N$=50$, made up of five parts, namely $[s_{1/2}\times p_{1/2}]^{(1)}$, 
$[s_{1/2}\times p_{3/2}]^{(1)}$, $[p_{1/2}\times d_{3/2}]^{(1)}$, $[p_{3/2}\times d_{3/2}]^{(1)}$ and 
$[p_{3/2}\times d_{5/2}]^{(1)}$ , we diagonalize the pairing matrix and obtain eigenvalues, that are all unbound, 
and the corresponding eigenvectors. We did calculations for three different values of pairing strength G 
i.e. $0$, $100$ and $200$ (upper limit to get all states unbound).
From Fig. (\ref{Dip1}), it is clear that a total of $10$ different transitions are possible from 
initial $0^+$ ground state to the final $1^-$ state of $^6$He.
We have investigated the detailed structure of $E1$ (dipole) strength distribution from two perspectives, one is to fix the pairing strength and second is to study the role of different configurations.
\begin{table}[!t]
\centering
\caption{Total B(E1) with varying pairing strength.}
\label{BE1T}
\begin{tabular}{cc}
\hline
G     & \begin{tabular}[c]{@{}c@{}}Total B(E1)\\ $e^2 fm^2$\end{tabular} \\ \hline
$0$   & 1.8747                                                           \\
$100$ & 1.8736                                                           \\
$200$ & 1.8378                                                           \\ \hline
\end{tabular}
\end{table}
Fig. (\ref{Dip2}), shows the total dipole transition strength of $^6$He with different values of G and Table- (\ref{BE1T}) tabulates the total B(E1) strength in $e^2fm^2$ with pairing strength G. 
As it should, it remains practically constant. The shape and strength of our dipole response function are consistent with the previous calculations \cite{Mik, Aoy3, Desco, Lay}. 
As a result of the smoothing procedure, the curves in Fig. (\ref{Dip2}) show a few minor wiggles, that are not to be attributed to resonances, but must be considered as an artifact. It is clear, though, that
there is an accumulation of strength at energies of $2-10$ MeV and possibly a shallow maximum around $3-5$ MeV.
We find in these calculations that the transition from $[p_{3/2}\times p_{3/2}]^{(0)}$ $\rightarrow$ $[p_{3/2}\times d_{5/2}]^{(1)}$ plays the dominant role in total dipole transition strength, 
whereas all the remaining nine transitions are less significant.

\section{Octupole strength distribution}
\begin{figure}[!t]
\begin{center}
\includegraphics[width=0.60\textwidth]{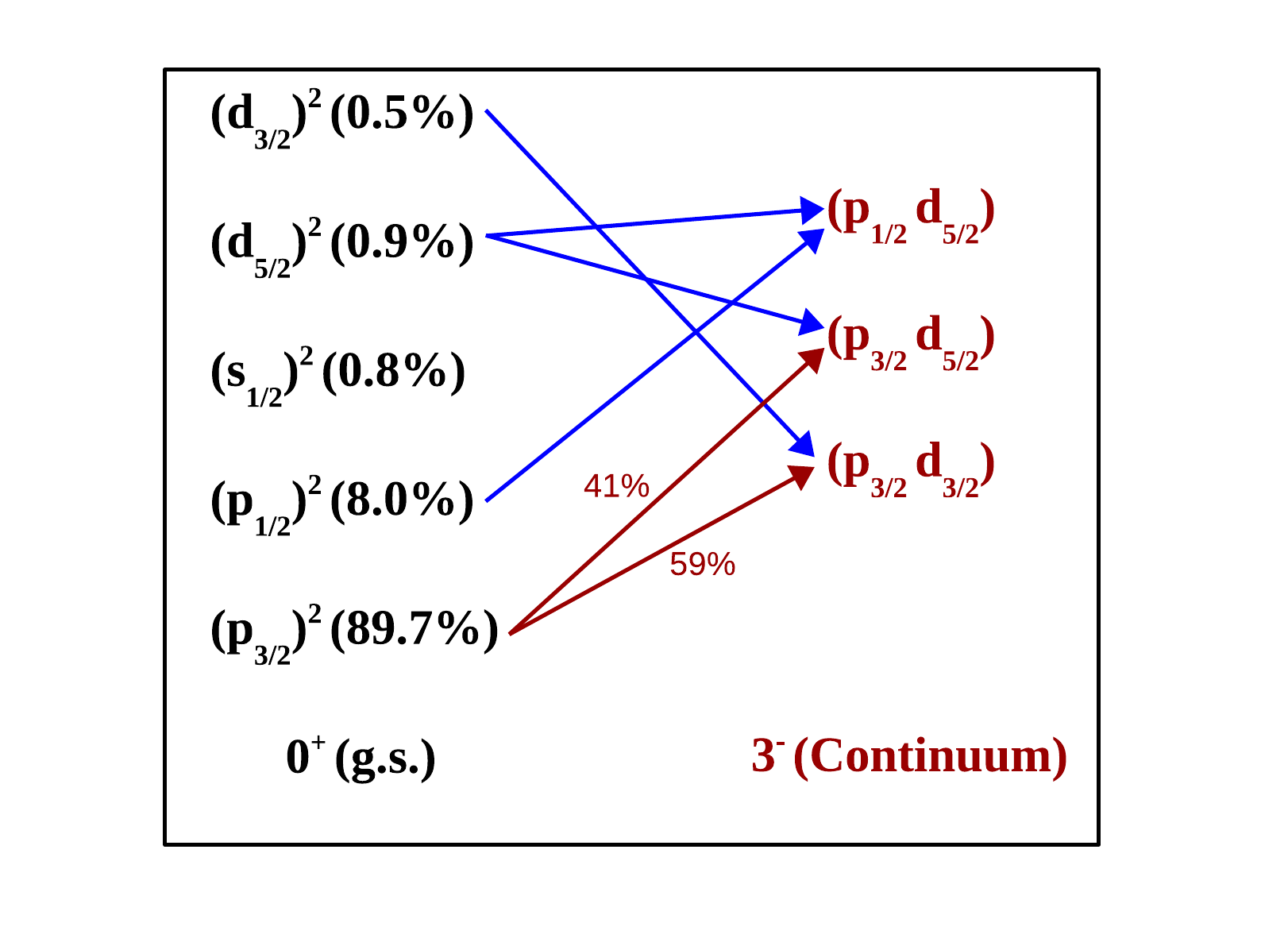}
\caption{(Color online) Schematic representation depicting all of the possible octupole transitions from ground state $0^+$ 
emerging from five different configurations to the final state $3^-$ emerging from three different 
configurations for $^6$He. The dominant transitions are highlighted in red color along with their percentage contribution.} 
\label{oct1}
\end{center}
\end{figure}

We have also investigated the detailed structure of $E3$ (Octupole) strength distribution of the system.
After constructing a basis of dimensions N$=50$, made up of three parts, namely 
$[p_{1/2}\times d_{5/2}]^{(3)}$, $[p_{3/2}\times d_{3/2}]^{(3)}$ and $[p_{3/2}\times d_{5/2}]^{(3)}$, 
we diagonalize the pairing matrix and obtain eigenvalues, that are all unbound, and the corresponding 
eigenvectors. We did calculations for four different values of pairing strength $G_3$ 
i.e. $0$, $250$, $500$ and $660$ (upper limit to get all states unbound).
From Fig. (\ref{oct1}), it is clear that there is a total of $6$ different transitions from 
initial $0^+$ ground state to the final continuum $3^-$ state of $^6$He.
\begin{table}[h]
\centering
\caption{Total B(E3) with varying pairing strength G.}
\label{BE3T}
\begin{tabular}{cc}
\hline
G     & \begin{tabular}[c]{@{}c@{}}Total B(E3)\\ $e^2 fm^6$\end{tabular} \\ \hline
$0$   & $91.2076$                                                        \\
$250$ & $91.1592$                                                        \\
$500$ & $91.0861$                                                        \\
$660$ & $90.8239$                                                        \\ \hline
\end{tabular}
\end{table}
\begin{figure}
\begin{center}
\includegraphics[width=0.70\textwidth]{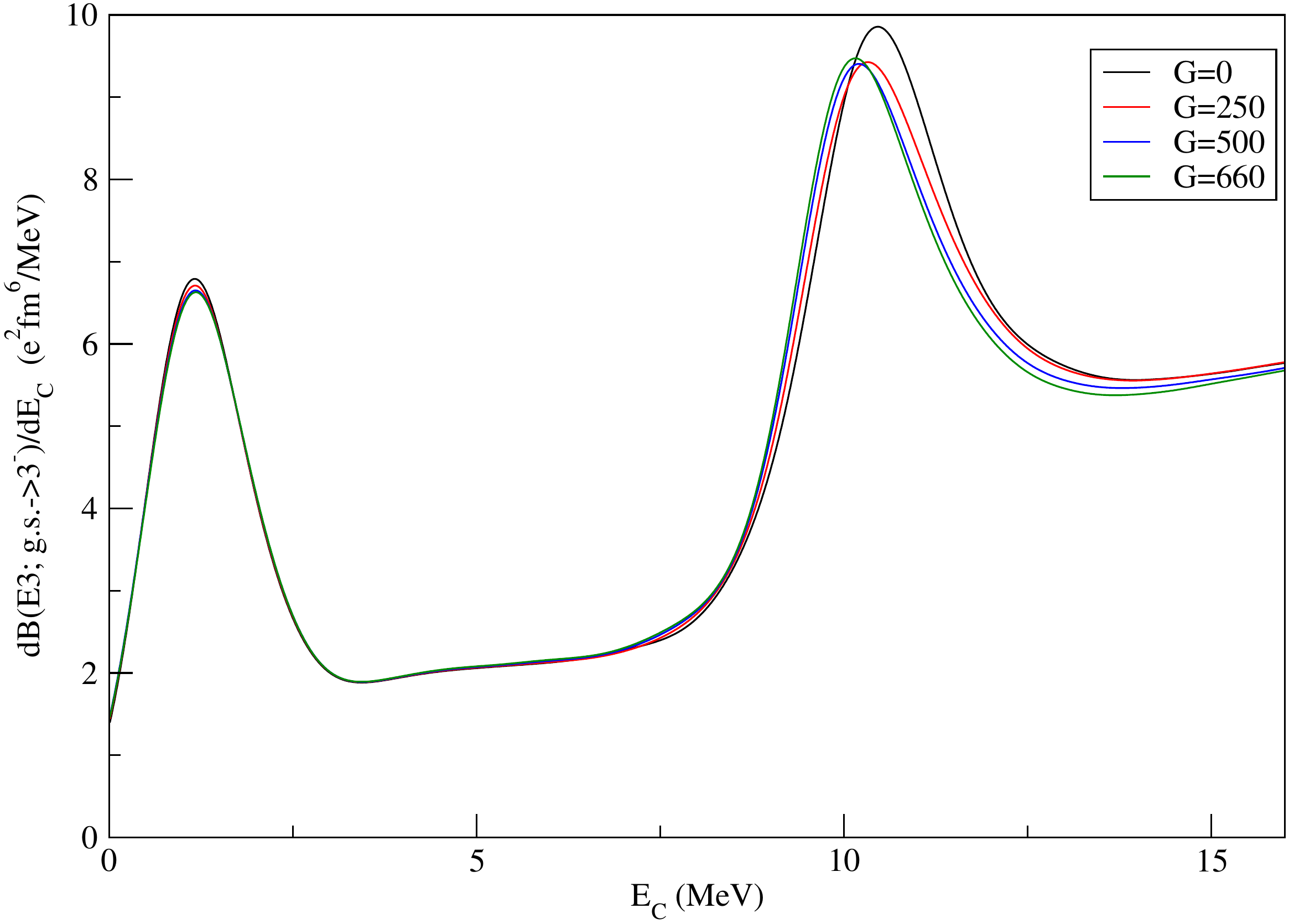}
\caption{(Color online) Octupole $E3$ transition strength distribution from ground state $0^+$ to the final state $3^-$ for $^6$He.} 
\label{oct2}
\end{center}
\end{figure}
We cannot integrate to find the total strength, because we can not extend the calculations beyond the present energy range due to computational limitations. Therefore it is not clear, at present, if we have reached 
the maximum value for the octupole distribution. Ideally, one should use a larger energy cut and maybe a smaller density of states. 
Table- (\ref{BE3T}) tabulates the total B(E3) strength in $e^2fm^6$ with pairing strength G up to the limit of $660$. We have estimated the total octupole strength to be approximately $190$ $e^2 fm^6$ by 
following the procedure outlined in Ref. \cite{hag}. Therefore our results exhaust about $50\%$ of the total expected strength.
Fig. (\ref{oct2}), shows the total octupole transition strength of $^6$He with different values of G.
The shape of our octupole response function clearly shows two large structures around $1$ MeV and $10$ MeV respectively, but the precise value of G has, in this case, little influence on the overall shape. 
This is due to the fact that with increasing \textit{l} the integral between different sets of single particle wave functions become progressively small and pairing becomes a weak perturbation.
We have found in these calculations that both these peaks take contribution from the transitions $[p_{3/2}\times p_{3/2}]^{(0)}$ $\rightarrow$ $[p_{3/2}\times d_{3/2}]^{(3)}$ 
and $[p_{3/2}\times p_{3/2}]^{(0)}$ $\rightarrow$ $[p_{3/2}\times d_{5/2}]^{(3)}$. These dominate the total octupole transition strength, amounting to approximately $\sim 59\%$ and $\sim 41\%$. 
All the remaining four transitions depicted in Fig. (\ref{oct1}) are comparatively less significant. 

\section{Conclusions}
\begin{figure}[!h]
\begin{center}
\includegraphics[width=0.75\textwidth]{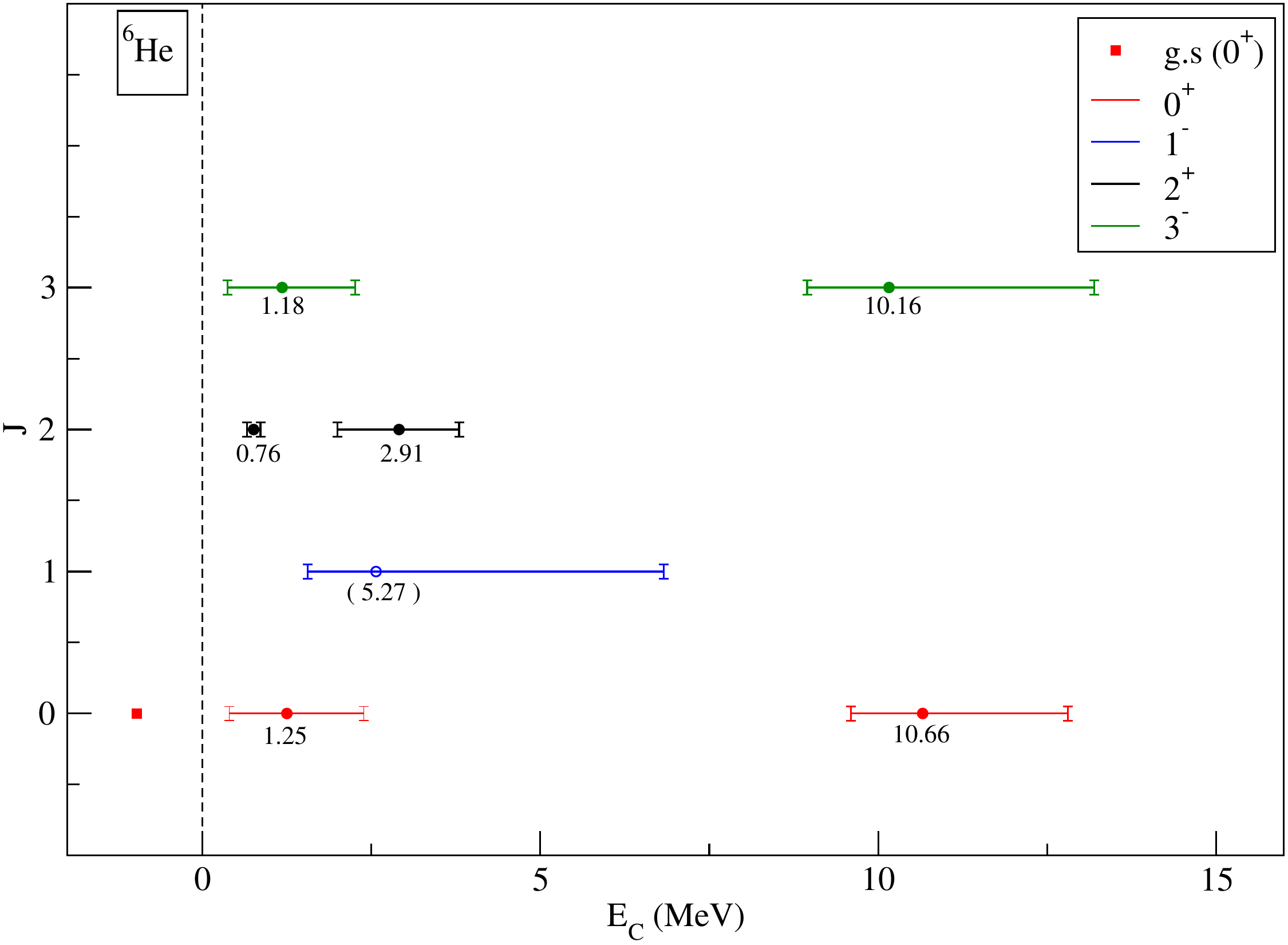}
\end{center}
\caption{(Color online) Schematic representation of the spectrum of $^6$He predicted by our simple model. The parenthesis in the J $=$ $1^-$ response indicates the uncertainty on the position of the peak (see text).}
\label{Res1}
\end{figure}
In summary, the electric multipole response of $^6$He has been investigated by using a simple structure model \cite{Fort, JS1, JS2, JST}, and the role of different configurations has been explored in each case. 
Fig. (\ref{Res1}), shows our predictions for the response of $^6$He to electromagnetic excitations of different multipolarity by showing the centroid of each state and the width on horizontal scale.
We have computed the B($E0$) values (Fig. \ref{Mon3}) from ground state to the continuum eigenstates and we have adjusted the strength of pairing matrix to get ground energy at right place. We have found two peaks at energies $1.25$ 
and $10.66$ MeV. Their widths are about $1.99$ and $3.21$ MeV respectively.
For dipole strength distribution (Fig. \ref{Dip2}) we have indicated in figure the case with maximal pairing strength that shows a maximum at $2.57$ MeV with asymmetric width of $5.27$ MeV.
For quadrupole strength distribution (Fig. \ref{oct2}) we have already reported in our previous calculation \cite{Fort} about the details of two resonances.
Finally for octupole strength distribution we have found two broad structures at $1.18$ and $10.16$ MeV with asymmetric widths of $1.89$ and $4.25$ MeV respectively. 
We expect that our efforts might be of help to unravel the complex patterns seen in the continuum spectrum of $^6$He.

\section{Acknowledgements}
We would like to thank J.A.Lay, K. Hagino, Sukhjeet Singh and Antonio Moro for useful suggestions. J.Singh gratefully acknowledges the financial support from Fondazione Cassa di Risparmio di Padova e Rovigo (CARIPARO).

%
%

\begin{thebibliography}{000}
%
%
\bibitem{Tanh1} I. Tanihata, H. Hamagaki, O. Hashimoto, Y. Shida, N. Yoshikawa, K. Sugimoto, O. Yamakawa, T. Kobayashi, and N. Takahashi, Phys. Rev. Lett. {\bf55}, (1985) 2676.
\bibitem{Tanh2} I. Tanihata et al., Phys. Lett. {\bf B 160}, (1985) 380.
\bibitem{Tanh3} I. Tanihata et al., Phys. Lett. {\bf B 206}, (1985) 592.
\bibitem{Tank} K. Tanaka et al., Phys. Rev. Lett. {\bf104}, (2010) 062701.
\bibitem{Zhu}  M.V. Zhukov, B.V. Danilin, D.V. Fedorov, J.M. Bang, I.J. Thompson and J.S. Vaagen,Phys. Rep. {\bf231}, (1993) 151.
\bibitem{Mat} M. Matsuo, Phys. Rev. {\bf C 73}, (2006) 044309.
\bibitem{Ber}  G.F. Bertsch, H. Esbensen, Ann. Phys. (N.Y.) {\bf209}, (1991) 327.
\bibitem{Ost} T. Otsuka, R. Fujimoto, Y. Utsuno, B. A. Brown, M. Honma and T. Mizusaki Phys. Rev. Lett. {\bf87}, (2001) 082502.
\bibitem{Kob} T. Kobayashi et al., Phys. Rev. Lett. {\bf60}, (1988) 2599.
\bibitem{Fuk} N. Fukuda et al., Phys. Rev. {\bf C 70}, (2004) 054606.
\bibitem{Aum} T. Aumann et al., Phys. Rev. {\bf C 59}, (1999) 1252.
\bibitem{Nak} T. Nakamura et al., Phys. Rev. Lett. {\bf96}, (2006) 252502.
\bibitem{Vitt4} F. Catara, C.H. Dasso and A. Vitturi, Nucl. Phys. {\bf A602}, (1996) 181.
\bibitem{Dan} B.V. Danilin, et al., Phys. Rev. {\bf C 55}, (1997) 577.
\bibitem{Brady} F.P. Brady et al., J.Phys.G {\bf10} (1984) 363.
\bibitem{Ajzen} F. Ajzenberg-Selove, Nucl. Phys. {\bf A 490}, (1988) 1.
\bibitem{Janeke} J. J\"{a}necke et al., Phys.Rev. C {\bf 54}, (1996) 1070.
\bibitem{Nakay} S. Nakayama et al., Phys. Rev. Lett. {\bf 85}, (2000) 262.
\bibitem{Nakay2} S. Nakayama et al., Phys. Rev. Lett. {\bf 87}, (2001) 122502.
\bibitem{Naka} T. Nakamura et al., Phys. Lett. {\bf B 493}, (2000) 209.
\bibitem{Naka2} T. Nakamura, Eur. Phys. J. A {\bf 13} (2002) 33-37.
\bibitem{Akimune} H. Akimune et al., Phys. Rev. C {\bf 67}, (2003) 051302\textbf{(R)}.
\bibitem{Moug} X. Mougeot et al., Phys. Lett. B {\bf 718} (2012) 441-446.
\bibitem{Povo} O.M. Povoroznyk, V.S. Vasilevsky, Ukr. J. Phys. {\bf 60}, (2015) 3.
\bibitem{Aoy} S. Aoyama, S. Mukai, K. Kato and K. Ikeda, Prog. Theor. Phys. {\bf93}, (1995) 99.
\bibitem{Aoy2} S. Aoyama, S. Mukai, K. Kato and K. Ikeda, Prog. Theor. Phys. {\bf94}, (1995) 343.
\bibitem{Myo1} T. Myo, S. Aoyama, K. Kato and K. Ikeda, Phys. Rev. {\bf C63}, (2001) 054313.
\bibitem{Cso} A. Csoto, Phys. Rev. {\bf C48},  (1993) 165.
\bibitem{Ara} K. Arai, Y. Suzuki and R. G. Lovas, Phys. Rev. {\bf C59}, (1999) 1432.
\bibitem{Mic1} N. Michel, W. Nazarewicz, M. Ploszajczak, K. Bennaceur, Phys. Rev. Lett., {\bf89}, (2002) 042502.
\bibitem{Mic2} N. Michel, W. Nazarewicz, M. Ploszajczak, K. Bennaceur, Phys. Rev. C {\bf 67}, (2003) 054311.
\bibitem{Hage} G. Hagen, M. Hjorth-Jensen, J.S. Vaagen, Phys. Rev. {\bf C 71}, (2005) 044314.
\bibitem{Mic3} N. Michel, W. Nazarewicz, M. Ploszajczak, Phys. Rev. {\bf C 82}, (2010) 044315.
\bibitem{Volya} A. Volya, V. Zelevinsky, Phys. Rev. Lett., {\bf94}, (2005) 052501.
\bibitem{Myo2} T. Myo, K. Kato and K. Ikeda, Phys. Rev. {\bf C 76}, (2007) 054309.
\bibitem{Zhu} M. V. Zhukov, B. V. Danilin, D. V. Fedorov, J. M. Bang, I. J. Thompson and J. S. Vaagen, Phys. Rep. {\bf 231}, (1993)  151-199.
\bibitem{Piep} S.C. Pieper, R.B. Wiringa, and J. Carlson, Phys. Rev. {\bf C 70}, (2004) 054325.
\bibitem{Nav1} P. Navr\'atil and W. E. Ormand, Phys. Rev. {\bf C 68}, (2003) 034305.
\bibitem{Nav2} P. Navr\'atil, J. P. Vary, W. E. Ormand, and B. R. Barrett, Phys. Rev. Lett. 87, (2001) 172502.
\bibitem{Dan1} B. V. Danilin, T. Rogde, S. N. Ershov, H. Heiberg-Andersen, J. S. Vaagen, I. J. Thompson, and M. V. Zhukov,Phys. Rev. {\bf C 55}, (1997) R577.
\bibitem{Dan2} B. V. Danilin, I. J. Thompson, J. S. Vaagen, and M. V.Zhukov, Nucl. Phys. {\bf A632}, (1998) 383.
\bibitem{Kor} S. Korennov and P. Descouvemont, Nucl. Phys. {\bf A740}, (2004) 249.
\bibitem{Dam} A. Damman and P. Descouvemont, Phys. Rev. {\bf C 80}, (2009) 044310.
\bibitem{Red} C. Romero-Redondo, S. Quaglioni, P. Navr\'atil and G. Hupin, Phys.Rev.Lett. {\bf  113}, (2014) 032503.
\bibitem{Mik} D. Mikami, W. Horiuchi, and Y. Suzuki, Phys.Rev. {\bf C 89}, (2014) 064303.
\bibitem{Fort} L.Fortunato, R.Chatterjee, Jagjit Singh and A.Vitturi, Phys. Rev. {\bf  90}, (2014) 064301.
\bibitem{JS1} Jagjit Singh, AIP Conf. Proc. {\bf 1681}, (2015) 020009.
\bibitem{JS2} Jagjit Singh, L.Fortunato, Acta Physica Polonica \textbf{B 47}, 1001 (2016).
\bibitem{JST} Jagjit Singh, Ph.D. thesis, Univ. of Padova, Italy (2016).
\bibitem{TUNL} TUNL, Nuclear Data Evaluation, \url{http://www.tunl.duke.edu/NuclData/General_Tables/5he.shtml}
\bibitem{HAG} K. Hagino and H. Sagawa, Phys. Rev. {\bf C 72}, (2005) 044321.
\bibitem{MYO} T.Myo et al., Progress in Particle and Nuclear Physics {\bf 79}, (2014) 1-56.
\bibitem{Kant} J. Kantele, Nucl. Instr. Meth. {\bf A271} (1988) 625.
\bibitem{Mey} J. Meyer, P. Quentin and M. Brack, Physics Letters {\bf 133B}, (1983) 279.
\bibitem{Aoy3} S. Aoyama, S. Mukai, K. Kato and K. Ikeda, Prog. Theor. Phys. {\bf116}, (2006) 1-35.
\bibitem{Desco} P. Descouvemont, E. Pinilla, and D. Baye, Prog. Theor. Phys. Suppl. {196}, (2012) 1.
\bibitem{Lay} J. A. Lay, A. M. Moro, J. M. Arias, and J. G\'{o}mez-Camacho, Phys. Rev. {\bf C 82}, (2010) 024605.
\bibitem{hag} H. Sagawa, N. Takigawa, Nguyen van Giai, Nucl. Phys. {\bf A543}, (1992) 575.

\end{thebibliography}
%

\end{document}